\documentclass[sigconf]{acmart}

\usepackage{xcolor}
\usepackage{graphicx}
\usepackage{multirow}
\usepackage{array}
\usepackage{caption}
\usepackage{color,soul}
\usepackage{hyperref}

\setcopyright{none}
\renewcommand\footnotetextcopyrightpermission[1]{}

\usepackage{fancyhdr}
\fancypagestyle{plain}{%
   \fancyhf{} %
   \fancyfoot[L]{ACM SIGCOMM Computer Communication Review}%
   \fancyfoot[R]{Volume 53 Issue 2, April 2023}%
}

\fancypagestyle{firstpagestyle}{%
   \fancyhf{} %
   \fancyfoot[L]{ACM SIGCOMM Computer Communication Review}%
   \fancyfoot[R]{Volume 53 Issue 2, April 2023}%
}

\begin{teaserfigure}
	\parbox{\textwidth}{\centering\normalsize
		This article is an editorial note submitted to CCR. It has NOT been peer-reviewed.\\
		The authors take full responsibility for this article's
		technical content. Comments can be posted through CCR Online.
	}
	\vspace{10pt}
\end{teaserfigure}

\usepackage{balance}
\begin{document}

\title{Recent Trends on Privacy-Preserving Technologies under Standardization at the IETF}

\author{Pratyush Dikshit}
\affiliation{
    \institution{CISPA Helmholtz Center for Information Security, \textit{DE}}
    \country{}
}
\email{pratyush.dikshit@cispa.de}

\author{Jayasree Sengupta}
\affiliation{
    \institution{CISPA Helmholtz Center for Information Security, \textit{DE}}
    \country{}
}
\email{jayasree.sengupta@cispa.de}


\author{Vaibhav Bajpai}
\affiliation{
    \institution{CISPA Helmholtz Center for Information Security, \textit{DE}}
    \country{}
}
\email{bajpai@cispa.de}

\begin{abstract}
End-users are concerned about protecting the privacy of their sensitive personal data that are generated while working on information systems. This extends to both the data they actively provide including personal identification in exchange for products and services as well as its related metadata such as unnecessary access to their location. This is when certain privacy-preserving technologies come into a place where Internet Engineering Task Force (IETF) plays a major role in incorporating such technologies at the fundamental level. Thus, this paper offers an overview of the privacy-preserving mechanisms for layer 3 (i.e. IP) and above that are currently under standardization at the IETF. This includes encrypted DNS at layer 5 classified as  DNS-over-TLS (DoT), DNS-over-HTTPS (DoH), and DNS-over-QUIC (DoQ) where the underlying technologies like QUIC belong to layer 4. Followed by that, we discuss Privacy Pass Protocol and its application in generating Private Access Tokens and Passkeys to replace passwords for authentication at the application layer (i.e. end-user devices). Lastly, to protect user privacy at the IP level, Private Relays and MASQUE are discussed. This aims to make designers, implementers, and users of the Internet aware of privacy-related design choices. 
\end{abstract}



\maketitle
\pagestyle{plain}

\section{Introduction}

The ever-increasing privacy risks accompanied by the growing concern amongst users have led different Internet bodies to focus on privacy preservation across all layers of the Internet architecture. With the establishment of the IETF, a mission over RFC 3935 \cite{IETF_RFC3935_Mission} was published focusing on a better human society through the influence of the Internet in communication, economics,
and education. As highlighted in RFC 8890, IETF is engaged in prioritizing the end-users by expanding the engagement of the Internet community, creating user-focused systems, and identifying and analyzing the negative impacts on end-users, etc. In order to retain its mission, IETF has undergone various technological advancements over the years to prevent unnecessary leakage of users' private information. In order to protect the privacy of individuals from unwanted tracking by location-tracking accessories, IETF recently discussed the best practices and protocols for accessory manufacturers whose products have built-in location-tracking capabilities in \cite{IETF_Location_tracker}.


\begin{figure}[!t]
    \centering
    \includegraphics[width=\columnwidth]{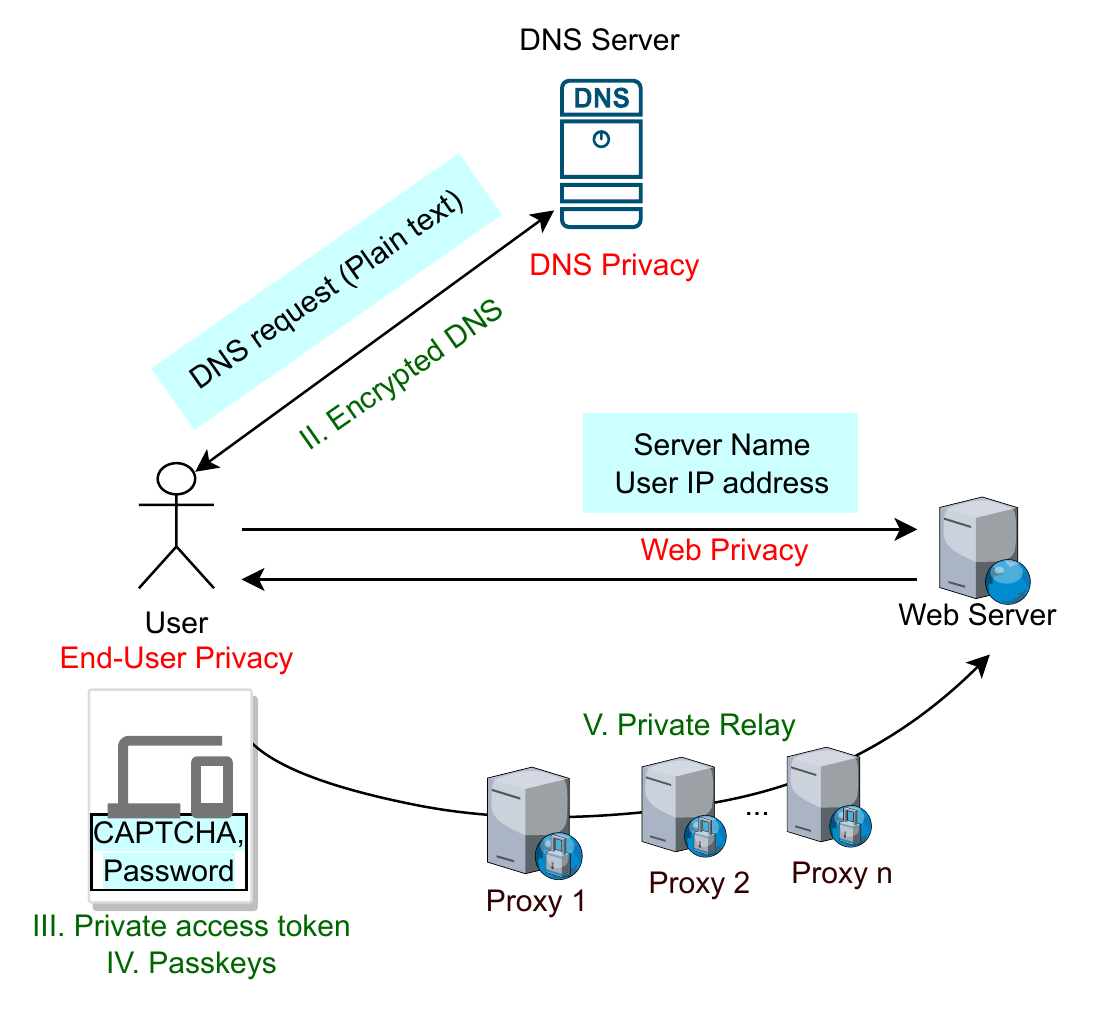}
    \caption{\em Privacy challenges arise during a typical communication between a user and the server. The challenges are highlighted in red while the corresponding privacy-preserving technologies are represented in green.}
    \label{fig:Image9}
    \vspace{-1em}
\end{figure}

\begin{figure*}[!t]
    \centering
    \includegraphics[width=\textwidth]{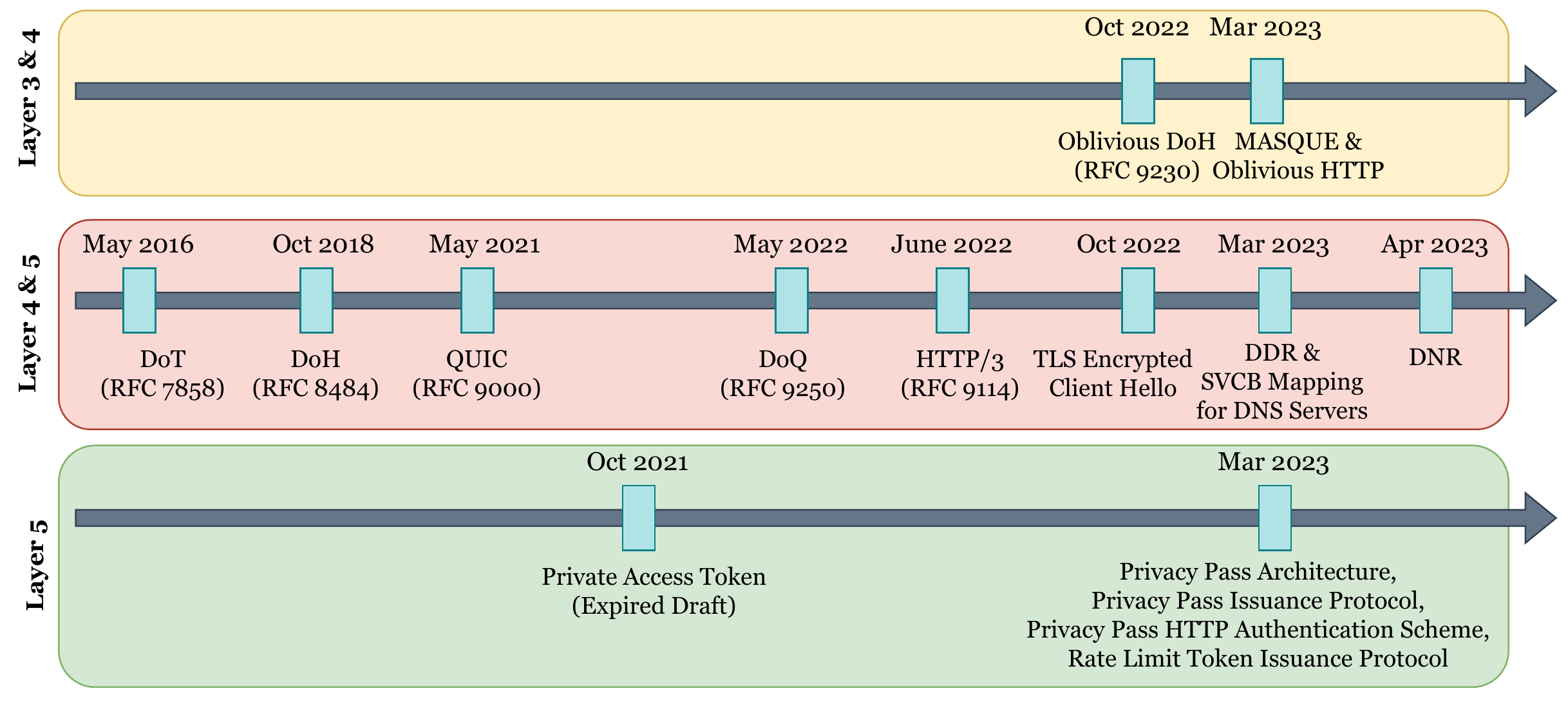}
    \caption{\em A timeline representing the standardization of privacy-preserving technologies at the IETF across layer 3 and above.}
    \label{fig:Image1}
    \vspace{-1em}
\end{figure*}

Privacy is the right of a person to manage the distribution of his/her personal information. Often such sensitive information may be prone to misuse. For example, traffic analysis, (e.g., correlation or measuring packet sizes), subverting the cryptographic keys used to secure protocols and the presence of active or passive wiretaps can lead to intrusion gathering at a large scale in the device/ network. These gatherings develop the trait to do widespread surveillance which is called Pervasive Monitoring (PM) and is declared as an attack in RFC 7258 \cite{IETF_PM}. With the surging rate at which data is being collected and digitally stored, it can lead to several privacy vulnerabilities. Out of those, the three major privacy concerns are privacy issues during DNS queries, end-user privacy risks on personal devices, and privacy breaches while communicating with the server through the Web. Thus, this paper highlights the privacy risks on the client's device as well as during communication between the client and the server. It then discusses the latest technologies that can mitigate PM attacks for layer 3 and above which are also currently under standardization at the IETF.

Let us consider the Web infrastructure as shown in Figure \ref {fig:Image9} where the user sends a DNS resolution request to the DNS server in plain-text \cite{DNS_encryption_survey}. Intruders (e.g. Internet Service Provider (ISP)) can view such requests and details of the requester which raises concerns about DNS Privacy. Encrypted DNS using TLS \cite{Houser_DoT}, HTTPs \cite{Csikor_DoH_Privacy} and QUIC \cite{QUIC_Kosek&Vaibhav} protocols described in RFC 9000 \cite{QUIC_Langley} act as a solution in this case and are broadly discussed in §\ref{sec:dns_privacy}. After fetching the IP address of the Web server from the DNS resolver, the user tries to communicate with the website through a browser. The website asks the user to authenticate itself using a Completely Automated Public Turing Test to Tell Computers and Humans Apart (CAPTCHA) challenge. The user also has to type in the password onto the browser which goes to the server in obfuscated form. Despite being obfuscated, the password still needs to be secure as it can lead to privacy leakage. Such privacy issues can be resolved by replacing the CAPTCHA with a token known as Private Access Token which is generated using the Privacy Pass Protocol. Details on the Privacy Pass protocol along with Private Access Tokens are available in §\ref{sec:PPP_PAT}. The replacement of password denoted as Passkey \cite{Passkeys_attack} (refer §\ref{sec:passkeys}) was first developed and implemented by Apple, Inc. Finally, the request packet from the user to the Web server contains the server name and the user IP address which is another privacy threat to the user as explained in RFC 6973 \cite{Privacy_consideration_rfc6973}. The proposal of deploying proxy servers between the user and the Web server referred to as Private Relay \cite{sattler_privaterelays} acts as a viable solution in this case. Hence, Private Relay and its IETF counterpart named MASQUE is described in §\ref{sec:PR_masque}. §\ref{sec: Discussion} presents open and future research directions in privacy-preserving technologies. §\ref{sec:conclusion} concludes the paper.

McQuistin \emph{et al.} \cite{McQuistin_IMC'21} explored IETF through the lens of RFCs and pointed out that the author pool is not quite diverse and is lacking representation from certain regions, especially South America and Africa. Keeping this in mind, the article aims to provide a lucid overview of the various privacy-preserving technologies under standardization at the IETF. This is mainly done to attract the audience who are not quite actively involved with the IETF so as to make them aware of the current ongoing standardization at the IETF in terms of designing privacy-preserving technologies for the Web. Figure \ref{fig:Image1} offers an overview of how IETF is progressing in its research in this domain, especially in recent times. Correspondingly, Table \ref{Table1: My tab} lists the respective documents and the working groups involved in developing those. We believe this article will inspire fresh minds to join the community and work on interesting privacy challenges and strengthen the research further.

\begin{table*}[]
\centering
\caption{\em List of IETF documents focussed on Privacy Protocols discussed in this paper.}
\resizebox{\textwidth}{!}{%
\begin{tabular}{@{}llll@{}}
\toprule

\textbf{WG} & \textbf{Document} & \textbf{Type} & \textbf{Reference} \\ \midrule
      &
Private Access Tokens &
Expired I-D &
\texttt{draft-private-access-tokens \cite{private-access-tokens-01}} 
\\

\hline

    TLS &
    TLS Encrypted Client Hello &
    ACTIVE I-D &
    \texttt{draft-ietf-tls-esni \cite{IETF_TLS_encrypted_Hello}} \\

\hline
   ADD   &
   Discovery of Designated Resolvers &

  ACTIVE I-D &
    \texttt{draft-ietf-add-ddr \cite{ietf-add-ddr-10}} \\

      &
   DHCP and Router Advertisement Options for the \\ & discovery of network-designated resolvers (DNR) &

  ACTIVE I-D &
    \texttt{draft-ietf-add-dnr \cite{ietf-add-dnr-14}} \\

        &
   Service Binding Mapping for DNS Servers &

  ACTIVE I-D &
    \texttt{draft-ietf-add-svcb-dns \cite{ietf-add-svcb-dns-08}}
    
    \\

\hline
Privacy & The Privacy Pass Architecture &
ACTIVE I-D &
\texttt{draft-ietf-privacypass-architecture \cite{ietf_privacy_pass_arch}} \\
Pass &
Privacy Pass Issuance Protocol &
ACTIVE I-D &
\texttt{draft-ietf-privacypass-protocol \cite{IETF_PP_Issuance}}
\\
&
Privacy Pass HTTP Authentication Scheme & ACTIVE I-D & \texttt{draft-ietf-privacypass-auth-scheme \cite{ietf-privacypass-auth-scheme-09}}
\\
&
Rate Limit Token Issuance Protocol & ACTIVE I-D & \texttt{draft-ietf-privacypass-rate-limit-tokens \cite{ietf-privacypass-rate-limit-tokens-01}}
\\
\hline
    
MASQUE &
IP Proxying Support for HTTP &
ACTIVE I-D &
\texttt{draft-ietf-masque-connect-ip \cite{ietf-masque-connect-ip-08}} \\ \bottomrule

\end{tabular}
}
\label{Table1: My tab}
\end{table*}

\section{DNS Privacy \label{sec:dns_privacy}}

With the growing privacy awareness in recent times, Web users are inclined towards preserving their data privacy \cite{tls.adoption}. As every communication on the Internet is preceded by a DNS lookup, the absence of traffic-level protection can lead to surveillance, censorship, manipulation, and rerouting \cite{Sandra_Encrypted_DNS}. Various research demonstrates that a user can be tracked over multiple websites by tracking his DNS queries \cite{DNS_USer_Session}\cite{Centralised_DNS}. Moreover, DNS traffic can also be used to collect information about IoT devices present in smart homes which eventually expose the utilization pattern of such devices as explained in \cite{IoT_Attack}\cite{IoT_Privacy_vulnarability}\cite{Policy_Privacy_IoT}. Currently, third parties such as ISPs can see every DNS request by the user. 

Existing technologies like TOR and VPN can resolve this issue up to a certain extent. For example, TOR encrypts the DNS request message that comes out from the TOR network's exit node, making the requester anonymous. Despite such advantages, a few limitations restrict the practical usage of TOR and VPN. As such TOR does not provide end-to-end encryption, nor does it encrypt the data before the entry node or after the exit node of the TOR network. However, VPN overcomes this problem by providing standard end-to-end encryption. Unlike TOR, VPN does not provide anonymity to its users. Moreover, security is limited to the TOR/ VPN application only. Hence, combining TOR and VPN is a suggested solution that comes at the cost of high latency. Thus, the inevitable problem of preserving privacy in DNS queries without compromising on speed still remains unresolved. This motivated the IETF to look into DNS privacy more seriously which resulted in the standardization of DoT (RFC 7858) \cite{TLS_rfc7858}, DoH (RFC 8484) \cite{ietf_RFC_DoH}, and DoQ (RFC 9250) \cite{RFC9250Dedicated_QUIC} recently. All of these protocols encrypt DNS communications between the stub resolver and the recursive DNS server thereby preserving user privacy.


\subsection{DNS-over-TLS (DoT)} 

In 2018, the IETF published its first RFC 8310 describing an approach to initiate TLS for DNS by using a dedicated DNS-over-TLS port. Encryption provided by TLS would then eliminate opportunities for eavesdropping on DNS queries in the network as highlighted in RFC 7258 \cite{IETF_PM}. However, this method requires a separate mechanism to upgrade a DNS-over-TCP connection \cite{mike:ccr:2022, Internet_resilience_IFIP23} on the standard port (TCP/53) to a DNS-over-TLS connection. Thus, with further improvements, IETF proposed a simplified protocol named DNS-over-TLS (DoT) in RFC 7858 \cite{TLS_rfc7858} which uses the well-known port 853 to specify TLS use thereby omitting the upgrade approach. DoT utilizes TLS to encrypt messages which helps it to differentiate DoT-based DNS requests from unencrypted DNS requests. Moreover, DoT traffic is also easy to differentiate from normal traffic by just filtering the port number. This makes DoT more secure in terms of greater control by the admin/ firewall on traffic flows. On the other hand, it makes DoT less legit in terms of preserving the privacy as 'type' of traffic (DNS queries) can be easily identified.

TLS 1.2 suffered from repeated security weakness which was addressed by regular updates and security patches as discussed by Bhargavan \emph{et al.} in \cite{TLS_Attacks_Bhargavan}. To upgrade TLS 1.2, it was necessary to improve the key exchange protocol during the handshake. Krawczyk \emph{et al.} \cite{TLS1.3_Krawczyk} highlights this improved TLS version (TLS 1.3) which includes a "0-RTT option" that is capable of reducing latency in cases where the client has previously retrieved or cached the public key of the server. This occurs by allowing the client to transmit protected information already in the first flow of the protocol. TLS 1.3 also provides perfect forward secrecy (PFS) in all modes by using the asymmetric key exchange. TLS 1.2 and TCP both require at least one round trip for the TCP handshake before the TLS handshake can happen. On the contrary, TLS 1.3 can establish the connection with 0-Round Trip Time (RTT). Lee \emph{et al.} \cite{TLS1.3_LeeKK21} discusses that the TLS 1.3 adoption rate is significantly faster than the previous versions of TLS. It took only 264 days for TLS 1.3 to be deployed by more than 15\% of websites after IETF officially approved the protocol. Comparatively, TLS 1.2 took around five years to achieve the same adoption rate. The adoption of TLS 1.3 is comparatively faster than its previous versions, but the research also highlights that 19.1\% of the TLS 1.3 adopted websites support TLS 1.3 unstably which may weaken the website’s security. The study further points out some limitations in the implementation of TLS 1.3 such as 0.03\% of the TLS 1.3 websites do not support downgrade protection, and many TLS libraries do not implement the parsing module for certificate extension fields essential for certificate-related extensions. These limitations affect the DNS-over-TLS implementation on a large scale as well. A proper standard configuration of TLS 1.3 is needed for stable support. Further studies are needed to understand the gaps between the existing TLS version and upgrading to TLS 1.3.

Currently, IETF is working on challenges for DNS clients to use DNS records to discover a resolver's encrypted DNS configuration. This is being done by the 'Discovery of Designated Resolver' (DDR) mechanism (see Table \ref{Table1: My tab}). DDR is introduced to securely discover parameters necessary for speaking to the same resolver using an encrypted transport in the presence of an active attacker that can add or modify packets, given the IP address of a DNS recursive resolver. In addition, IETF is also looking into the SVCB mapping for named DNS servers, allowing them to indicate support for encrypted transport protocols (see Table \ref{Table1: My tab}). DoT has the potential to increase the volume of network traffic, which may have implications for network performance and scalability. 

\subsection{DNS-over-HTTPS (DoH)} 

The IETF had a few early proposals on end-to-end communication over HTTPS which lacked agreement on what the appropriate formatting should be where they didn't follow HTTP best practices. Later, IETF standardized DNS-over-HTTPS (DoH) as a protocol in RFC 8484 \cite{ietf_RFC_DoH} for sending DNS queries and getting DNS responses over HTTPS where each DNS query-response pair is mapped into an HTTP exchange. DoH uses port 443 which is the same as the usual HTTP connection. TLS and HTTPS protocols are configured and send out GET/ POST requests along with a path extension- “url/dns-request?dns=NAME”. The Web server needs to take care of differentiating between the HTTP request and the DNS request. RFC 8484 \cite{ietf_RFC_DoH} further explains the integration of DNS protocol with HTTP that to provide a transport medium suitable for both existing DNS clients and native Web applications requesting access to the DNS. While it is easy to block DoT traffic using port-based filtering 
\cite{Csikor_DoH_Privacy}\cite{Houser_DoT}, it is infeasible to do the same for DoH traffic as it would also block normal Web traffic.

DoH leads to a secure DNS traffic flow and third parties proxy an encrypted DNS request. Only users and Google/ Cloudflare (in this case), but not third parties, can look into the DNS traffic. Secondly, DoH prevents DNS Spoofing. Though DoH prevents eavesdroppers from directly reading the content of DNS messages, clients cannot send the request to the DNS server and receive the response without revealing their local IP address, thereby, revealing their identity. To counter this, IETF proposed Oblivious DoH (ODoH) (RFC 9230) \cite{IETF_9230_ODoH} which is an enhanced mechanism to hide the client's IP address while sending the DNS query. ODoH is built on top of DoH that allows proxied resolution, in which the DNS message is encrypted and no single server is aware of either the IP or the message. However, service providers such as Google/ Cloudflare can still gather IP addresses at scale (if not message content) which leads to a data centralization problem~\cite{doan:toit:2022, doan:ifipnw:2021}. Further, eliminating content filtering is unfavorable to a few countries borders. Another limitation of DoH was discovered by Siby \emph{et al.} \cite{Sandra_Encrypted_DNS} by applying a machine learning approach. The study shows how an adversary can extract traffic features and train the model to analyze DNS activities based on tuples like packet size, the timing between two packets, TLS headers, and the direction of the traffic. Currently, there is limited support for DoH which can lead to a fallback attack in which a malicious actor creates fake negotiations (e.g., through man-in-the-middle) between the client and server. This leads to a less-secure protocol being used in their communication \cite{DNS_encryption_survey}. 


\subsection{DNS-over-QUIC (DoQ)} 

QUIC was developed by Google \cite{QUIC_Langley} and later standardized by the IETF in RFC 9000 \cite{RFC9000QUIC_Basic, mike:commag:2021} to improve Web performance \cite{tanya:tnsm:2022}. As TCP, QUIC is connection-oriented, and thus multiple channels can be created between the client and the server where channel/ path selection is independent of the service provider's selected path. However, QUIC is encapsulated in the UDP datagram which leads to faster deployment in user space, unlike the kernel space as in the case of TCP. QUIC merges the transport (TCP) and cryptographic (TLS) handshakes into one and is built on top of TLS 1.3, thus supporting 0-RTT as well \cite{WebPrivacy_Design_IFIP23}. Hence, QUIC is at least 1-RTT faster than TCP with TLS. Keeping these features of QUIC in mind, in 2020 the Network Working Group of IETF proposed a draft stating the use of QUIC to provide transport privacy for DNS. They also classified the usage of DNS protocol into three groups: stub to the recursive resolver, recursive resolver to an authoritative server, and server to server. But the initial design only focused on the "stub to recursive resolver" scenario. Later, IETF improved the protocol and published RFC 9250 which standardized the usage of DNS-over-QUIC (DoQ) \cite{RFC9250Dedicated_QUIC} for all three scenarios.

Recently, IETF standardized HTTP version 3 (HTTP/3) \cite{DOH3_measurement} in RFC 9114 \cite{IETF_rfc9114} which provides transport for HTTP semantics having an internal framing layer similar to HTTP/2 while using QUIC as the underlying transport protocol. During implementation, when a client learns that HTTP/3 is available on a server, it opens up the QUIC connection. QUIC then negotiates with the client to set up a communication channel. HTTP/2 relies on in-order transmission of packets which QUIC does not guarantee, but HTTP/3 uses separate unidirectional streams to modify and track field table states. Hence, DNS-over-HTTP/3 (DoH3) \cite{IETF_rfc9114}assures the in-order transmission of packets. At present, Cloudflare, Google, etc. have successfully implemented DoH3 on Firefox and Android.

The limitation of DoQ arises with TLS 1.3 (RFC 8446) which is not supported by most of the Web platforms at present as discussed in \cite{QUIC_Kosek&Vaibhav}. As a result, the QUIC protocol cannot run over certain carriers and providers. Moreover, HTTP/3 implementation is limited to a specified maximum size for the message header. If a server receives a larger header packet, then it may discard the packet with an HTTP 431 error message. Further, DoQ requires the use of the QUIC protocol, which may not be supported by all network infrastructure devices. This means that there may be compatibility issues when trying to implement DoQ in some networks. Therefore, further research is needed to identify potential compatibility issues and to develop solutions to address them. On the other hand, the first \textit{'Hello'} message is sent unencrypted in QUIC and thus also in DoQ as mentioned in RFC 9250. This raises several security vulnerabilities as attackers are able to manipulate these packets. To counter such malicious activities, recently the TLS Working Group of the IETF proposed a draft implementing Encrypted Client Hello (ECH) \cite{IETF_TLS_encrypted_Hello} (see Table \ref{Table1: My tab}) where the very first \textit{'Hello'} message is also encrypted under the server's public key. This work is still in progress and awaiting security analysis. 

\section{Privacy Pass Protocol \& Private Access Tokens \label{sec:PPP_PAT}}

The Web services such as emails, search engines, and APIs over the Internet are continuously experiencing threats from bots. Distinguishing the humans from the bot was addressed by CAPTCHA. The CAPTCHAs are indeed proven mechanisms for providing anti-fraud protection on many websites but are also tedious. Thus, CAPTCHA hinders the user experience and makes the websites less accessible. Though CAPTCHA is primarily intended to distinguish humans from bots, Nakatsuka \emph{et al.} \cite{CAPTCHA_NakatsukaOPT21} and Motoyama \emph{et al.} \cite{CAPTCHA_MotoyamaLKMVS10} show that CAPTCHAs are not very effective in doing this task. Many CAPTCHAs can be solved by algorithms (e.g., image recognition software) or outsourced to human-driven CAPTCHA farms to be solved on behalf of bots. This is the reason why privacy-minded folks at Fastly, Inc. developed Private Access Token (PAT) to replace CAPTCHA as an alternative for some supported platforms such as iOS. PATs provide much broader privacy benefits at the client level than the CAPTCHA does. PATs offer greater control over access to sensitive data or systems. With CAPTCHA, once the challenge is completed, the user is typically granted access to the system or data. In contrast, private access tokens can be used to grant access to specific resources or systems, with the option to revoke access at any time. PATs offer better compliance with privacy regulations. CAPTCHA may collect and process personal data such as images or audio recordings, which can pose privacy concerns for users. In contrast, private access tokens do not require any personal data to be collected or processed, making them more compliant with data privacy regulations. PAT uses Public Key Cryptography (PKC) where the authorization process does not require human interaction. These tokens are developed by using Privacy Pass Protocol.

\begin{figure}[!t]
    \centering
    \includegraphics[width=\columnwidth]{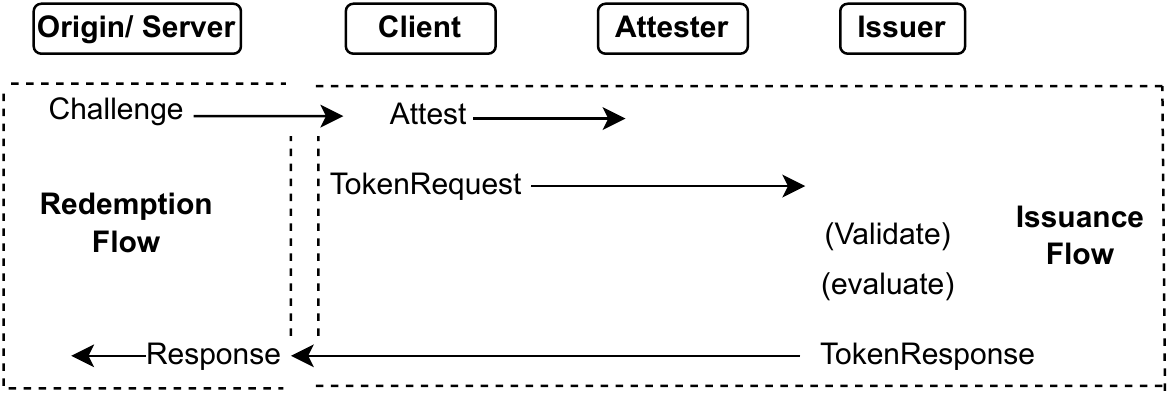}
    \caption{\small\textit{ Architecture of Privacy Pass Protocol: The challenge is solved in Issuance Flow between the Client, the Attester, and the Issuer which will eventually be sent back to Origin/ server as a Response in Redemption Flow}}
    \label{fig:Image3}
    \vspace{-1em}
\end{figure}
  
The Privacy Pass protocol was introduced by the IETF Charter Privacy Pass WG (see Table \ref{Table1: My tab}) for authorizing clients that had already been authorized in the past without compromising their privacy. The Privacy Pass Protocol interacts with the websites for anonymous user authentication. In certain time-sensitive communications, it only needs to check if the user is already authorized for the same service without collecting any further information about the user. For example, let's consider a client attempting to authenticate itself to a server. The authorization mechanism does not reveal to the server anything about the client. In a nutshell, a user solves a random Internet challenge once for which she gets some randomly generated nonce values (Tokens) by the Privacy Pass which are cryptographically blinded (values are not feasibly linkable) and then sent to the Origin/server. If the response is found correct, then the provider signs the blinded token and returns them to the client. The privacy pass then unbinds the tokens, called PAT, and stores them for future use. The tokens are "privacy-preserving" in the sense that they cannot be linked back to the previous session where they were issued as proposed by the Network Working Group of IETF (refer Table \ref{Table1: My tab}).

Privacy Pass is composed of two protocols: Issuance and Redemption. In an architectural view, the issuance protocol runs between a client and two network functions: Attestation and Issuance (refer Table \ref{Table1: My tab}) as shown in Figure \ref{fig:Image3}. The Issuer is trusted by the server/ origin and is responsible for issuing tokens in response to requests from the client. The Attester is responsible for attesting properties of the client. The Issuance protocol is a 2-message protocol that takes as input a challenge from the redemption protocol and produces a token. Recent development took place in 2022 where the privacy pass protocol includes a 2-message issuance provision of privately verifiable and publicly verifiable tokens as explained in the IETF draft of Issuance Protocol (see Table \ref{Table1: My tab}). The significance of having a publicly verifiable issuance protocol is that any Origin can select a given Issuer to produce tokens, as long as the Origin has the Issuer's public key, without explicit coordination from the Issuer. 

Privacy Pass is a challenge-response mechanism implemented by Content Delivery Networks (CDNs) that focuses on reducing the influx of malicious requests. As a result, if a CDN does not implement such security measures then the utility of Privacy Pass would be limited \cite{PrivacyPass_ALex_Popoet}. Moreover, due to the limited support of DoH, Google, and Cloudflare are implementing this protocol on their browsers first. The initial version of the Privacy pass protocol on server-side support is implemented by Cloudflare, Inc. and the corresponding client-side implementation already exists as Chrome and Firefox browser extensions. The privacy pass working group at IETF has an open challenge of setting up a ceiling on the number of tokens issued to a client in a single issuance phase.  If there is no limit, malicious clients could abuse this and cause excessive computation which may lead to a Denial-of-Service attack as suggested by PrivacyPass WG.

\begin{figure}[!t]
    \centering
    \includegraphics[width=\columnwidth]{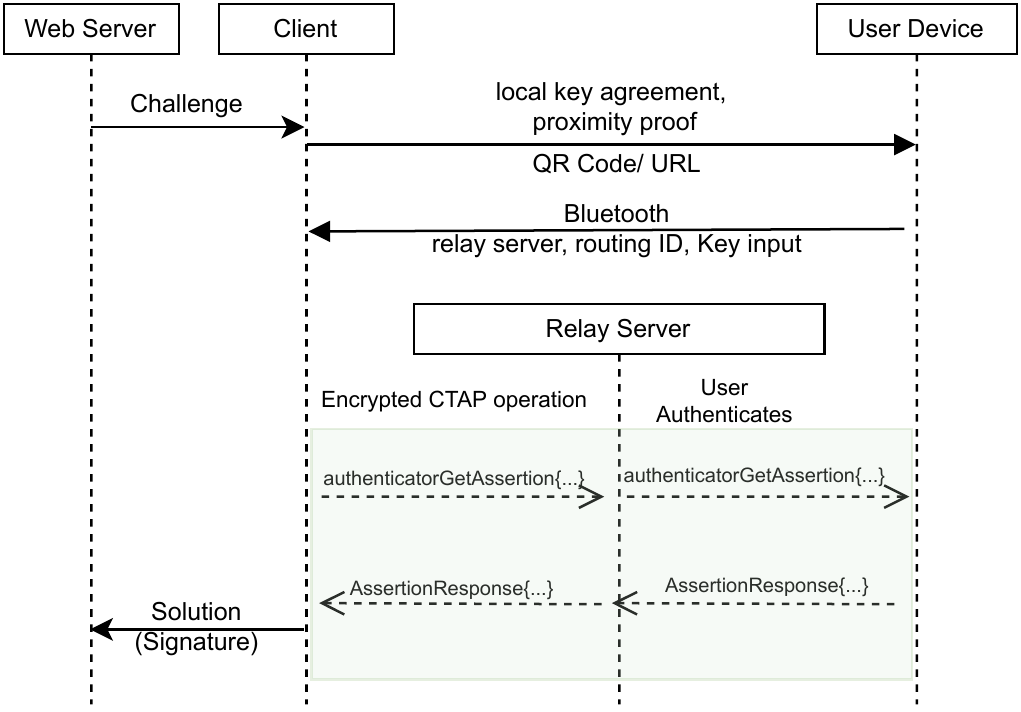}
    \caption{\small\textit{Passkeys Based Authentication. A client is capable of signing in to another device in local proximity using Bluetooth and authenticating with the help of a relay server. The web server is not at all involved in the signing process except by providing challenges and validating the signature.}}
    \label{fig:Image7}
    \vspace{-1em}
\end{figure}

\section{Passkeys \label{sec:passkeys}}

The traditional form of authentication requires a username and password. But this method poses certain security risks. Especially, phishing attacks can lead to the credentials getting exposed. This is why 'Security Key' \cite{Auth_Security_Keys} was introduced as a device that facilitates access and strong authentication into other devices and applications and is also referred to as Security Token. These tokens are generally easy to use with a single or no-click authentication. These are based upon Web Authentication API (WebAuthn) \cite{WebAuthn} standard which is more secure than traditional passwords. WebAuthn (Web Authentication) is a standard for password-less authentication that uses biometrics, such as fingerprints or facial recognition, or physical security tokens, such as USB keys. WebAuthn is designed to work with web browsers and online services, making it a convenient and secure method for users to authenticate without the need for passwords. The WebAuthn allows servers to register and authenticate using PKC instead of a password.

Eventually, researchers realized that authentication could be done seamlessly without any extra device/ authenticator. Thereby, researchers at Apple, Inc. introduced Passkeys which are a type of credential that use cryptographic tools to leverage biometric features (e.g. Touch-ID, face-recognition). Passkeys act as a replacement for the traditional password-based authentication that leads to improved user privacy. Passkeys are built on WebAuthn and new flexible UI options. Passkeys are a type of password-less authentication technology that relies on public key cryptography. In addition, the Passkeys have inherited security features of WebAuthn \cite{WebAuthn}. When creating a passkey, a unique key-pair is generated which works only for the application it is created for. The public key is saved on the server and the private key stays on the device itself. The user sends a sign-in request to the server. The server sends a single-use challenge to the device as shown in Figure\ref {fig:Image7}. This challenge is based on one of the WebAuthn algorithms (e.g. EAS 256 problem). The device produces a digital signature using the user's private key and then sends it back to the server. The server validates the signature using the respective public key. If it is validated, the user gets logged in. Through this mechanism, the server is confirmed that the user has the private key without learning any information about the key. Thus, Passkeys cannot be threatened by phishing and is thereby less attractive to the attacker.

Another interesting feature about Passkeys is that the same user account can be logged in from other devices as well without sharing the Passkeys. The client contains a QR code that has a single-use encryption key. The authentication device produces a Bluetooth advertisement containing the routing information for a relay server as shown in Figure \ref{fig:Image7}. This local exchange allows selecting a server through which routing information is shared. It has two major advantages. It performs out of band key agreement which the server can’t see. Secondly, the two devices are in physical proximity, hence, a remote attacker cannot have Bluetooth advertisement details. The relay server cannot learn any secret information as the devices follow standard encrypted Client To Authenticator Protocol (CTAP) operations \cite{Auth_Security_Keys}. For example, Account Manager creates Passkeys in iOS which are stored in the iCloud Chain that can be used in any Apple device authenticated by the same user. While signing into another device using Bluetooth, there are a few attacks possible such as Man-in-the-middle attack, sniffing, jamming, packet-injection, session hijacking, etc. \cite{Passkeys_attack}. This is inherited from the limitations of Bluetooth technology. Moreover, there is a scope for improvement in Passkeys Recovery. If the device is lost and the Passkeys are not linked with the cloud, the recovery procedure would adopt a traditional method that can be improved. Passkeys are initially launched by Apple, but others are close behind. For example, Microsoft will likely launch its own equivalent soon, although it may not initially be compatible with Apple’s implementation. Chrome on Windows stores passkeys in Windows Hello, which doesn't synchronize them to other devices as of May 2023, Chrome on macOS stores passkeys in a local profile and doesn't synchronize them to other devices, Passkeys from iCloud Keychain isn't yet available in Chrome on macOS.

Though Passkeys are typically based on WebAuth algorithms, there are a few fundamental differences between them with respect to their implementations. One of the key differences between passkeys and WebAuthn is the type of authentication method used. Passkeys use public key cryptography, while WebAuthn relies on biometrics or physical security tokens. Additionally, passkeys require the user to generate a key pair and manage their private key, while WebAuthn does not require the user to manage any cryptographic keys. Another difference between the two technologies is their scope of use. Passkeys are typically used for specific applications or services that require secure authentication, while WebAuthn is designed for use across the web and can be used for a wide range of online services and applications. The question that arises here is \textit{"Which authentication is better than the other?"}. The answer relies on the purpose of implementing the authentication method. Considering the client-side privacy on mobile devices, passkeys provide better flexibility as passkeys can be used for a wide range of applications and services, while WebAuthn is primarily designed for use with web browsers and online services. Summarily, we can conclude by saying that Passkeys have the potential to make passwords obsolete, and social media platforms are well-positioned to adopt passkeys at an early stage of deployment. Passkeys could have plausible applications in bio-metric authentication of employees in an organization, or membership of a forum/ conference/ community where the count of people is more important than that of people’s identity.

\begin{figure}[!t]
    \centering
    \includegraphics[width=\columnwidth]{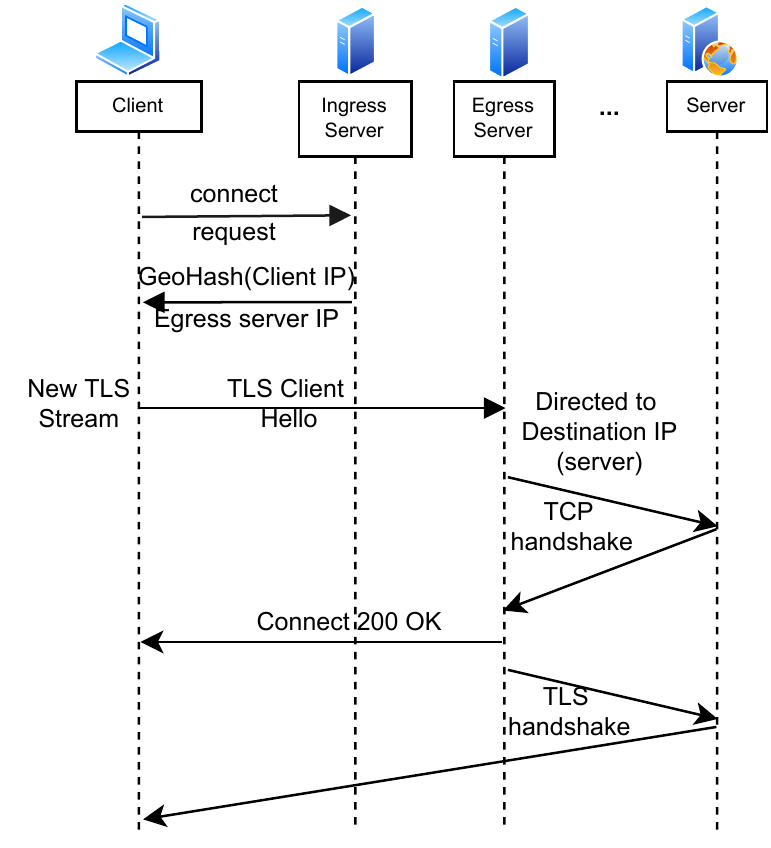}
    \caption{\small\textit{Communication through Private Relays. The client connects with the server using its GeoHash IP address through Proxy Relay Servers. Only the ingress can identify the client whereas only the egress knows the targeted server.}}
    \label{fig:Image5}
    \vspace{-1em}
\end{figure}

\section{Private Relays and MASQUE \label{sec:PR_masque}}

Each HTTP request to a server reveals certain information about the client via the Source IP Address. Connection reuse aggravates the problems as servers can correlate requests that share a connection. As mentioned earlier in §\ref{sec:dns_privacy}, VPN and Tor can be beneficial in such circumstances. However, using such techniques degrades the overall performance as every request needs a new TLS connection. This is where certain privacy-enhancing technologies proposed by the IETF like Oblivious HTTP, Private Relays along with Masques come into the picture to help solve the problem.

Private relay is a privacy-enhancing mechanism applied when communicating with the cloud by establishing proxy server(s) between them. Private Relay claims to protect the user's privacy while browsing, seeking DNS queries, and running insecure HTTP traffic. Private relay separates client's IPs from servers and uses multi-hop MASQUE (Multiplexed Application Substrate over QUIC Encryption) proxy as described by MASQUE Working Group, IETF (refer Table \ref{Table1: My tab}) for proxying IP packet, whereas oblivious DoH \cite{IETF_9230_ODoH} (RFC 9230) is used for other traffic. MASQUE proposes the use of QUIC as a substrate protocol, allowing arbitrary data to be tunneled over QUIC. Quite interestingly, Oblivious DoH makes sure that the server does not get any information about the IP address of the client as well as the message content. Proxies are authenticated with TLS 1.3 raw public keys and clients are authenticated with RSA blind signatures. On the contrary, Oblivious HTTP \cite{ietf_draft_oblivious_http} offers to send encrypted HTTP messages from a client to a gateway via a trusted relay service. IETF claims to use a combination of encapsulation and relaying to enable connection reuse. It also ensures that nobody sees the client's IP address as well as the plaintext HTTP message remains hidden from the relay resources.

Recently, iOS 15 and macOS implemented the private relay for all Safari browser traffic, all DNS traffic, all unencrypted HTTP traffic, and Mail Pixel Trackers. The main goal of developing a private relay is that not a single entity in the communication channel should have visibility of both - the user's IP address and what the user is accessing. To design this mechanism, developers marked two hops as a minimum requirement for the separation of connection data. Here, two hops signify the two proxy servers. Clients select the hops and have nested encryption for handshakes to the next hop. The hops are chosen to be run by separate proxies. But, collusion would be required to track user activities across the communication channel. Currently, this is handled by designing policies for not allowing data sharing. Also, in order to understand the website/ server compatibility in Private Relay, rough geolocation is preserved, when the user wants it. The egress proxy gets the hint of the geohash of the client and an appropriate egress IP/ Proxy2 is selected. Proxy1 (Ingress proxy) knows the actual IP address of the client but does not know the destination address. The client requests the address of Proxy2. Proxy1 provides the GeoHash of the client's IP address and the IP address of Proxy2. The Client communicates with the Proxy2 using its GeoHash provided by Proxy1. Here, Proxy2 can see the destination address of the client and directs the client to the destination (cloud). This entire mechanism is depicted in Figure \ref{fig:Image5}. Apple currently provides an implementation for private relays on MacOS and iOS via a paid subscription. 

Sattler \emph{et al.} \cite{sattler_privaterelays} performed measurements to detect clients communicating via Apple's Private Relay Network which described that iCloud Private Relay influences the operation of passive network analysis, and due to its workings, it can also impact intrusion detection systems (IDSs). Moreover, the ingress addresses can be used to identify relay traffic as a passive network observer. K{\"{u}}hlewind \emph{et al.} \cite{Mirja_MASQUE} performed an evaluation of MASQUE-based Proxying and suggested that for mobile networks, this might provide an opportunity to accept network configurations with simpler link layer loss recovery schemes and only use local loss recovery when explicitly required by an application.

\section{Future Research Directions \label{sec: Discussion}}

In spite of the faster adoption of TLS 1.3, around 20\% of TLS 1.3 adopted websites have unstable support which may weaken the security and results in privacy breach in DNS queries over TLS. In this context, the research direction to look at the limitations and reasons behind the lack of stable TLS 1.3 support would be interesting. 
A small proportion of websites (0.03\%) of all the TLS 1.3 websites do not support downgrade protection. It would be interesting to find out the cost at which the websites do not support the downgrade protection and what alternative protection can be provided in such cases. 
Moreover, DoT can be integrated with other security protocols, such as DNSSEC where DNSSEC provides data integrity and authentication for DNS responses, while DoT encrypts DNS traffic, protecting it from eavesdropping and tampering. However, it is yet to be explored from the viewpoint of the potential benefits and challenges of integrating DoT with such security protocols. 
The adoption rate of HTTP/3 \cite{HTTP3_adoption} is around 25\% of the top 10 million websites in the twelve months. The question is to evaluate the adoption trend \cite{encrypted_dns_adoption_trend} of DoH/3 empirically for all the DNS resolvers with respect to their geolocations could be interesting research. This could provide insight into the gaps where the adoption lacks. The adoption of QUIC protocol \cite{QUIC_adoption} is less than 8\% which gives room to further investigate why there is a depletion in the rate of adoption in the last one year from 9\% (approx.) to 7.3\%. The adoption of DoQ~\cite{mike:pam:2022} is even lesser. 
Thus, the question is to find out the reasons behind the reduction in the adoption rate would be an open avenue. This also gives an understanding of the limitations such as infrastructure, regulatory policies, etc. by which large chunks of resolvers do not adopt DoQ. Within this context, it would be further interesting to investigate the performance and behavior change of DoQ by varying the underlying OS platforms and programming languages. Furthermore, the performance variation of DoQ by using different mediums at the physical layer such as co-axial cables or fiber would be an interesting new direction. 
DDR has an attraction for the networking community to look into. One can exploit the SVCB record in order to verify if a DNS resolver supports DDR, and which security protocol is being implemented such as HTTP/2, HTTP/3, TLS, or QUIC by sending an SVCB query (type64) for the reserved name ${\_dns.resolver.arpa}$. This study is scalable over all the existing DNS resolvers which can provide the spread of DDR support in various geographical zones.

At present, the adoption of Private Access Token is scarce which strikes a question on what the possible reasons are, behind the hindrance of the adoption of PAT. One of the reasons is the unavailability of DoH. More research is needed to find out some other reasons as well. Secondly, in cases where token issuance experiences delays, a website could display an indicator to signify the waiting period or switch to an alternative method of user validation or fallback to the CAPTCHA method. Hence, the focus should be on how to evaluate this fallback scenario by analyzing all the websites that have adopted the private access token as an authentication mechanism. 
Also, the Privacy pass protocol could be integrated with more CAPTCHA techniques in order to enhance authentication while preserving user privacy \cite{PrivacyPass_ALex_Popoet}. With cryptographic tools 
the privacy pass protocol can be explored with quantum-safe verifiable oblivious pseudorandom function (VOPRF) \cite{VOPRF_JareckiKK14}. This protocol can be implemented in anonymous session resumption for TLS, anonymous referral code mechanism, and single-bit zero-knowledge proof.

Recently, Google introduced Passkey support on Android and Chrome \cite{passkeys_google} for Android 9 or higher which shows a major adoption. It would be interesting to have an insightful understanding of how the behavior of Passkeys varies across different combinations of Android versions and browsers. This may lead to figuring out an optimized methodology for integrating passkeys between two or more different platforms, operating systems, etc.  Considering the control management of passkeys, researchers can also work around limiting the optimized number of signing-in to the devices to reduce the risk in case of failure/ leakage. Passkeys are still in the process of enhancement for further research. 
Passkeys rely on public key cryptography, which requires support from both client and server ends. Further study is needed to explore ways to ensure interoperability between different passkey implementations and between different applications and services. Passkeys leverage cryptographic tools to utilize the biometric features that lead to further research to explore ways to make passkeys accessible to all users, regardless of their physical or cognitive abilities.

In the private relay, the introduction of a minimum of two proxy servers introduces a notable latency overhead, which becomes more pronounced as the number of proxy servers increases. So, evaluating the trade-off between the number of proxy servers and users' privacy is an invitational opportunity. Secondly, it would be interesting for the researchers to work around the optimized number of servers required on a case-by-case basis considering the cost of latency. For example, the number of proxy servers required in military mobile communication might be more than that of digital-payment-related communication. Thirdly, the analysis of how the communication between the client and the server behaves in case of an outage/ failure of one of the proxy servers could be an open research avenue. 

\section{Conclusion \label{sec:conclusion}}

This article offers a first impression of the recent privacy techniques addressed in IETF, and how their interplay extends user privacy challenges. Over the years, privacy-preserving techniques are evolving. With DNS encryption at the core of this evolution, its future development will be built on the foundation of DoH and DoQ, thereby extending its reach to increasingly more supporting platforms. Also, considering the issues with the secret key (or password) which is vulnerable to being stolen, misplaced, or compromised in one way or the other, this paper discussed the Passkeys technology. Passkeys are capable of replacing the secret key (passwords) with a challenge-response conversation method between the user device and the server.  While privacy at the user end has multiple variations, the private access token system promises to supersede traditional methods in order to prevent unnecessary data leakage. Several improvements have happened on the Web to separate the client's IP address from the origin, however, private relay and MASQUE using multi-hop architecture have finally accomplished the goal. This will further enhance the rejuvenation of the Web, thus aiding the development of next-generation Web applications.

\begin{acks}

This work was supported by the Volkswagenstiftung Niedersächsisches Vorab (Funding No. ZN3695). We thank Simone Ferlin \{\textit{Red Hat, Sweden}\} for her insightful suggestions and constructive comments during the course of preparing this article.

\end{acks}

{ \balance
{
\bibliographystyle{unsrt}
\bibliography{reference}
}
}

\end{document}